\documentclass[useAMS,usenatbib]{mn2e}
\usepackage{amsmath}
\usepackage{amssymb}
\usepackage{epsfig}
\usepackage{natbib}
\title[The filling factor of intergalactic metals at $z=3$]{The filling factor of intergalactic metals at redshift z=3}
\author[C. M. Booth et al.]{C. M. Booth,$^{1}$\thanks{E-mail: booth@strw.leidenuniv.nl} Joop Schaye,$^{1}$ J. D. Delgado$^{1}$ and Claudio Dalla Vecchia$^{1,2}$\\
$^{1}$Leiden Observatory, Leiden University, P.O. Box 9513, 2300 RA Leiden, The Netherlands\\
$^{2}$Max-Planck-Institut f\"ur Extraterrestrische Physik, Giessenbachstra\ss{}e, D-85478 Garching, Germany}
\newcommand{\ion}[2]{\hbox{#1\,{\sc #2}}}
\newcommand{\ionsubscript}[2]{\hbox{\scriptsize #1\,{\tiny #2}}}

\newcommand{\hi}{\ion{H}{I}}
\newcommand{\civ}{\ion{C}{IV}}
\newcommand{\ovi}{\ion{O}{VI}}
\newcommand{\hisub}{\ionsubscript{H}{I}}
\newcommand{\civsub}{\ionsubscript{C}{IV}}

\newcommand{\tauciv}{\tau\civsub}
\newcommand{\tauhi}{\tau\hisub}
\newcommand{\zenrich}{Z}
\newcommand{\renrich}{r_{\rm z}}
\newcommand{\menrich}{m_{\rm z}}

\newcommand{\msun}{{\rm M}_\odot}

\newcommand{\zsun}{{\rm Z}_\odot}
\newcommand{\fmass}{f_{\rm m}}
\newcommand{\fvol}{f_{\rm V}}

\begin{document}
\pagerange{\pageref{firstpage}--\pageref{lastpage}} \pubyear{2011}
\maketitle
\label{firstpage}

\begin{abstract}
  Observations of quasar absorption line systems reveal that the $z=3$
  intergalactic medium (IGM) is polluted by heavy elements down to
  $\hi$ optical depths $\tau_{\rm HI}\ll10$. What is not yet clear,
  however, is what fraction of the volume needs to be enriched by
  metals and whether it suffices to enrich only regions close to
  galaxies in order to reproduce the observations. We use gas density
  fields derived from large cosmological simulations, together with
  synthetic quasar spectra and imposed, model metal distributions to
  investigate what enrichment patterns can reproduce the observed
  median optical depth of $\civ$ as a function of $\tau_{\rm HI}$. Our
  models can only satisfy the observational constraints if the $z=3$
  IGM was primarily enriched by galaxies that reside in low-mass
  ($m_{\rm tot}<10^{10}\,\msun$) haloes that can eject metals out to
  distances $\ga 10^2$~kpc. Galaxies in more massive haloes cannot
  possibly account for the observations as they are too rare for their
  outflows to cover a sufficiently large fraction of the
  volume. Galaxies need to enrich gas out to distances that are much
  greater than the virial radii of their host haloes.  Assuming the
  metals to be well mixed on small scales, our modeling requires that
  the fractions of the simulated volume and baryonic mass that are
  polluted with metals are, respectively, $>10\%$ and $>50\%$ in order
  to match observations.
\end{abstract}


\section{Introduction}
\label{sec:intro}

Studies of quasar absorption spectra indicate that at redshift $z
\gtrsim 2$ most of the baryons in the Universe reside outside of
galaxies \citep[e.g.][]{rauc97,wein97,scha01} in the gas that is
observable through $\hi$ Ly$\alpha$ absorption and that pollution of
the low-density intergalactic medium (IGM) by heavy elements is
wide-spread
\citep[e.g.][]{tytl95,cowi95,cowi98,elli00,scha00,scha03,cars02,arac04,simc04,pier04,song05,agui08,pier10}.
The very fact that heavy elements are able to make their way from
galaxies out into intergalactic space, has far reaching implications
for the galaxy formation process.  We know that metals are only
produced in the high-density environments where star-formation occurs,
and in order to reach the low-density IGM, they must have been removed
from these regions.  Energetic feedback processes that drive gas out
of galaxies are believed to be the most important way in which the IGM
is polluted by metals, as suggested by both observational
\citep[e.g.][]{pett01,shap03,stei10} and theoretical
\citep[e.g.][]{agui01a,theu02,cen05,oppe06,torn10} studies.

In spite of the observational evidence that metals have polluted the
high-redshift IGM, several issues remain unclear. What fraction of the
volume of the IGM, and hence of the Universe, is polluted by metals?
Out to what distance do galaxies need to enrich the gas around them in
order to reproduce observations?  Can observed galaxies do the job or
are fainter galaxies the main culprit?

The metal distribution in the high-redshift IGM has been investigated
by many authors using self-consistent, hydrodynamical simulations
\citep{theu02,agui05,cen06,oppe06,koba07,oppe08,tesc10,shen10,wier09,wier10,cen10}. The
interpretation of these simulation results is, however, complicated by
their complex nature and by the crudeness of, and freedom provided by,
the required subgrid models.  In addition, their computational expense
prohibits comprehensive explorations of parameter
space. Observationally, \citet{pier04} use four quasar spectra along
with a suite of synthetic quasar spectra and find that the lower limit
for the volume filling factor of \ovi\, is $>4\%$.

Simple models therefore represent a useful complement to full-blown
simulation studies. To this end various authors have employed models
in which the IGM is enriched by spherical
\citep[e.g.][]{mada01,bert05,scan02,scan06,samu08} or anisotropic
\citep[e.g.][]{agui01a,agui01b,pier07,germ09,pins10} bubbles of metals
placed around haloes to investigate the metal distribution in the
universe. Models in which the gas is enriched down to varying densities have been used to constrain the volume filling factor of enriched gas \citep[e.g.][]{scha03,pier04}. However, as discussed by e.g.\ \citet{scha05}, the inferred filling factors could be misleading if the metals are poorly mixed, as obervations suggest to be the case on both large \citep{scha03} and small scales \citep{scha07}.

In the present work we combine toy models for the metal distribution
with a large, cosmological hydrodynamical simulation to investigate
what halo masses could host the sources of the observed intergalactic metals, out to what distances the galaxies need to enrich the gas, and what (large-scale) volume filling factor of heavy elements is required in order to
reproduce the observed metal distribution as probed through absorption
lines in the spectra of quasars.  In particular, we compare model
predictions to the observed $z=3$ relation between the median optical
depth of $\civ$ as a function of the $\hi$ optical depth
\citep{scha03}. We choose to restrict our analysis to the median
optical depth because it allows us to compare with published results
and because it provides a simple measure of how far away from galaxies the IGM is being enriched.

We achieve this by extracting synthetic absorption spectra after
imposing simple metal distributions in which all haloes above a given
mass cut are allowed to enrich the IGM spherically out to a fixed
radius.  This allows us to link a metal distribution with a
well-defined mass and volume filling factor, via cosmological gas
density and temperature distributions, to the observations and so to
determine which of the possible metal distributions are capable of
reproducing the observations.  We will show that the IGM must have
been primarily enriched by galaxies that reside in low-mass ($m_{\rm
  tot}<10^{10}\,\msun$) haloes and are capable of driving gas out to
distances $>10^2\,$kpc. Assuming the metals to be well mixed on small
scales, the fractions of the volume and baryonic mass that are
polluted with metals are, respectively, $>10\%$ and $>50\%$ in all
models that are capable of matching the observations.

This paper is organised as follows.  We first introduce our simulation set
(Sec.~\ref{sec:simulations}) and the methods we use to distribute the
metals (Sec.~\ref{sec:metals}), then, in Sec.~\ref{sec:results}, we
describe our results and in Sec.~\ref{sec:conclusions} we summarise
our findings and conclude.

\section{Method}

\subsection{Simulations}
\label{sec:simulations}
The simulation analysed in this study is one of the cosmological,
hydrodynamical simulations that comprise the OverWhelmingly Large
Simulations (OWLS) project and is described in detail in
\citet{scha10}.  Briefly, the simulation was run using a significantly
extended version of the parallel PMTree-Smoothed Particle
Hydrodynamics (SPH) code {\sc gadget iii} \citep[last described in
][]{spri05b}, a Lagrangian code used to calculate gravitational and
hydrodynamic forces on a particle by particle basis.  The simulations
track star formation, supernova feedback, radiative cooling and
chemodynamics, as described in \citet{scha08}, \citet{dall08},
\citet{wier08} and \citet{wier09}, respectively. This physical model
is denoted as the \emph{REF} model in \citet{scha10}, and is used in
all of the simulations analysed in this paper.

For the purposes of this work, our prescriptions for radiative cooling
and reionization are the most important aspects of the model, as the
thermal state of the IGM depends on them.  In brief, we calculate
radiative cooling and heating using the tables of \citet{wier08},
which contain net cooling rates \citep[calculated using the code {\sc
    cloudy}, last described in ][]{ferl98} as a function of density,
temperature and redshift for each of the 11 elements hydrogen, helium,
carbon, nitrogen, oxygen, neon, magnesium, silicon, sulphur, calcium
and iron, computed under the assumption of ionization equilibrium and
in the presence of the \citet{haar01} model for the uniform, evolving
meta-galactic UV and X-ray radiation field from galaxies and quasars
as well as the cosmic microwave background.  The simulations model
hydrogen reionization by switching on the \citet{haar01} background at
$z=9$. Helium reionization is modelled by heating the gas by a total
amount of 2 eV per atom. This heating takes place at $z=3.5$, with the
heating spread in redshift with a Gaussian filter with
$\sigma(z)=0.5$. This reionization prescription used in these
simulations matches the temperature history of the IGM inferred from
observations by \citet{scha00temp} \citep[see Fig.~1 of][]{wier09}.

\begin{table}
\caption{Simulation parameters. From left to right: Simulation
  identifier; comoving box size ($L_{\rm box}$); number of particles
of both dark matter and gas; mass of dark matter particles ($m_{\rm DM}$), and
initial mass of gas particles ($m_{\rm gas}$).}
\begin{center}
\begin{tabular}{r|r|r|r|r}
Simulation & $L_{{\rm box}}$ & $N$ & $m_{\rm DM}$ & $m_{\rm gas}$ \\
& (Mpc/$h$) & & ($\msun$) & ($\msun$) \\
\hline
\emph{L025N512}  & 25.0 & 512$^3$ & $8.68\times10^6$ & $1.85\times10^6$ \\
\emph{L025N256}  & 25.0 & 256$^3$ & $6.95\times10^7$ & $1.48\times10^7$ \\
\emph{L012N512}  & 12.5 & 512$^3$ & $1.09\times10^6$ & $2.31\times10^5$ \\
\emph{L012N256}  & 12.5 & 256$^3$ & $8.68\times10^6$ & $1.85\times10^6$ \\
\end{tabular}
\end{center}
\label{tab:sims}
\end{table}

The simulations used in this paper are summarised in
Table~\ref{tab:sims}.  All assume a flat $\Lambda$CDM cosmology with
the cosmological parameters:
$\{\Omega_{\rm m},\Omega_{\rm b},\Omega_\Lambda,\sigma_8,n_{\rm s},h\}=\{0.238,0.0418,0.762,0.74,0.951,0.73\}$,
as determined from the WMAP 3-year data \citep{sper07}.  The
simulations from which we derive the bulk of our results are
\emph{L025N512} and \emph{L012N512}. Two additional simulations,
\emph{L025N256} and \emph{L012N256}, are used to independently assess
the effects of simulation box size and numerical resolution.

Although each of the simulations was run to $z=2$ we restrict our
analysis to the $z=3$ simulation snapshots, approximately
corresponding to the median redshift of the observational sample that
we compare to.  Our analysis depends on the identification of the
masses and locations of gravitationally bound dark matter haloes,
which are identified using the spherical overdensity criterion
implemented in the {\sc SubFind} algorithm \citep{spri01}.  Halo
properties quoted in this paper are defined with respect to spheres
with radius $r_{200}$ and mass $m_{200}$, centred on the potential
minimum of each identified halo, defined so that they contain a mean
internal density equal to 200 times the critical density of the
Universe at the redshift we are considering. 

We note that although we will present results only for the reference
implementation of the subgrid physics modules, we have repeated the
analysis for a range of physics implementations.  This is important
because the different physics prescriptions can affect the density,
temperature and velocity fields of the absorbing gas, changing the
predicted ion abundances and optical depths.  We find, however, that
using either a simulation with strong, AGN feedback
(\emph{AGN\_L025N512} in the OWLS nomenclature), or a simulation that
neglects both supernova feedback and cooling through metal lines
(\emph{NOSN\_NOZCOOL\_L025N512} in the OWLS nomenclature) has a
negligible effect on our results or conclusions.

\subsection{Imposed metal distributions}
\label{sec:metals}
In order to predict a synthetic $\tauciv-\tauhi$ relation, we require
knowledge of both the distribution of metals and the physical state of
the absorbing gas.  In the simulation, gas metallicities are tracked
self-consistently, but in the present study we do not make use of this
information. We instead assume that all haloes with a total mass
greater than $\menrich$ are able to enrich the surrounding gas out to
a proper distance $\renrich$ to a metallicity $\zenrich$, and that
outside of these spheres the metallicity of the IGM is zero. Our model
for the intergalactic metal distribution is therefore completely
specified by three parameters: $\menrich$, $\renrich$ and
$\zenrich$. As we will see, the parameters $\renrich$ and $\menrich$
determine the shape of the relation between $\tauciv$ and $\tauhi$,
which is the primary focus of this paper.  The metallicity changes
only the normalisation, with $\tauciv\propto\zenrich$, so we simply
scale $\zenrich$ in each run to match the normalisation of the
observed $\tauciv-\tauhi$ relation at $\log_{10}(\tauhi)=2.5$, the
largest optical depth probed by the observations.

At this point we must note one caveat: we have imposed metal
distributions on to already completed simulations, the models are not
fully self-consistent in that they do not include the effect of the
winds that carry the metals on the density and temperature structure
of the gas and in that they do not include the effect of the metals on the cooling rates. The last point is, however, not a major concern as we will show that the metallicities required to match the
  observations are sufficiently low (typically $10^{-3}-10^{-2}Z_\odot$; Fig.~\ref{fig:od}), that metals do not
  significantly change the gas cooling rates \citep[e.g.][]{wier08}.

  While the hydrodynamical simulations that underlie our
models did include winds, these winds fall short of being able to
account for the observed $\civ$ at low $\tau_{\rm HI}$, as we will
show elsewhere. This failure is actually consistent with our
results. Although the simulations have sufficient resolution to
identify dark matter haloes to very low masses (corresponding to $\sim
10^2$ dark matter particles), their finite resolution does cause us to
strongly underestimate the star formation rates in most of the
low-mass haloes that we can identify. Hence, the simulations
underestimate the number and strength of the outflows originating from
the low-mass haloes that we claim to be responsible for the enrichment
of the IGM. Because not all of the gas that is enriched in our models
was touched by winds in the underlying hydro simulation, we cannot
exclude the possibility that a self-consistent simulation giving rise
to a similar distribution of bubbles would predict the enriched gas to
be too hot to be visible in $\civ$. Reassuringly, we find, as noted above, that post-processing simulations without winds or with much stronger winds leads to identical conclusions.

We note that we expect the heating effect of
outflows from low-mass galaxies to be smaller than those from the more
massive galaxies that our simulations do include. This is because
there are already many low-mass galaxies at high redshift, giving the
gas more time to cool, and because they are observed to drive winds of
velocities $\la 10^2$\,km/s \citep[e.g.][]{schw04,mart05} (this
velocity corresponds to post-shock temperatures of $\sim
2\times10^5$\,K, assuming the gas is fully ionized and of primordial
composition), which leaves the post-shock gas at temperatures for
which the cooling time is much shorter than the age of the Universe
\citep{wier08} .

The conclusions based on our simple models will, however, ultimately need to be
confirmed by self-consistent, hydrodynamical simulations. Unfortunately, at the
moment such simulations rely on uncertain subgrid models for
the generation of winds \citep[e.g.][]{dall08} and they lack the
resolution required to model outflows from low-mass galaxies and to
simulate the small-scale mixing relevant for the observations
\citep{scha07}. 

The bulk of our results are derived from a grid of models in which
$\menrich$ is varied in steps of 0.5~dex from the lowest mass haloes
that can be robustly identified in the highest resolution simulation,
$\menrich=10^8\,\msun$, which are the least massive haloes that are
expected to be able to produce stars after reionization
\citep{efst92,quin96,thou96}, up to $\menrich=10^{11}\,\msun$.

Note that a mass of $10^{11}\,\msun$ is small compared with the total
masses inferred for observed galaxies at $z=3$.  For example,
\citet{adel05} find that Lyman-break galaxies reside in haloes of mass
$\sim10^{12}\,\msun$.  However, as we will show, such high-mass
galaxies are unimportant for the enrichment of the IGM.

The parameter $\renrich$ is changed in factors of 2 from $31.25$~kpc
to 500~kpc.  In addition, we investigate a set of runs in which haloes
in the fiducial simulation are allowed to enrich the IGM out to a
fixed multiple of their virial radius, $r_{200}$.  Galactic winds with
velocities up to 400-600~km/s are frequently detected in starburst
galaxies through the gas absorption lines that are blue-shifted
relative to their host galaxies \citep[e.g.][]{pett01,stei10,raki10}.
If we assume that winds were ejected from galaxies at high redshift
($z\gg3$) and that their velocities do not decrease with time, then by
$z=3$ galaxies can enrich out to a maximum radius of $0.9-1.4$~proper
Mpc.  The assumption of a constant, high outflow velocity and launch
at $z\gg 3$ make this estimate far too optimistic, but all of the
models that match the observations require $\renrich$ to be no larger
than $500$~kpc (which is likely still too optimistic; see
e.g.\ \citealt[e.g.][]{agui01a}), and are thus compatible with
constraints placed on the metal distribution by travel-time arguments.

Optical depth distributions are calculated by firing $10^3$ randomly
chosen lines of sight through the simulation volume and calculating
absorption spectra for both $\hi$ and $\civ$ following the procedure
outlined in e.g. Appendix~A4 of \citet{theu98}.  The mean $\hi$
optical depth in our simulations at $z=3$ is $\tau_{\rm eff}=0.388$, which is consistent
with observations \citep[e.g.][]{scha03,fauc08}.  In order to match
observations with HIRES on the Keck telescope, we convolve our spectra
with a Gaussian line-spread function with a full-width-at-half-maximum
of 6.6 km/s, and resample our spectra onto 1 km/s pixels.  We do not
add noise to our spectra, but we have verified that the addition of
Gaussian noise with a signal-to-noise ratio of greater than 25 does
not significantly affect any of our results.  We generate absorption
spectra for two transitions: $\hi$ (1215.67\AA) and the $\civ$ doublet
(1548.20\AA, 1550.78\AA). In order to compare to the observed
$\tauciv-\tauhi$ distribution, we then bin pixels in $\tauhi$ and
calculate the median $\tauciv$ corresponding to the redshifts of the
pixels in each $\hi$ bin.

For each run, in addition to measuring the $\tauciv-\tauhi$ relation,
we calculate the fraction of the total mass ($\fmass$) and volume
($\fvol$) that has been enriched from
\begin{equation}
\fmass \approx \frac{\sum_i m_i(Z>0)}{\sum_i m_i}\,\,\,{\rm and}\,\,\,\fvol \approx
\frac{\sum_i h^3_i(Z>0)}{\sum_i h^3_i}\,.
\label{eq:frac}
\end{equation}
Here, $m_i$ and $h_i$ are the SPH particle mass and smoothing kernel,
respectively, and the sums in the numerator of each fraction extend
only over particles with non-zero metallicity.  We have verified that
using $m_i/\rho_i$ instead of $h_i^3$ gives nearly identical volume filling fractions, as expected. For the solar abundance\footnote{This
  corresponds to the value obtained using the default abundance set of
  {\sc CLOUDY} \citep[version 07.02; last described by][]{ferl98}.} we
use the metal mass fraction $\zsun=0.0127$.

\section{Results}
\label{sec:results}

\begin{figure*}
\begin{center}
\includegraphics[width=\textwidth,clip]{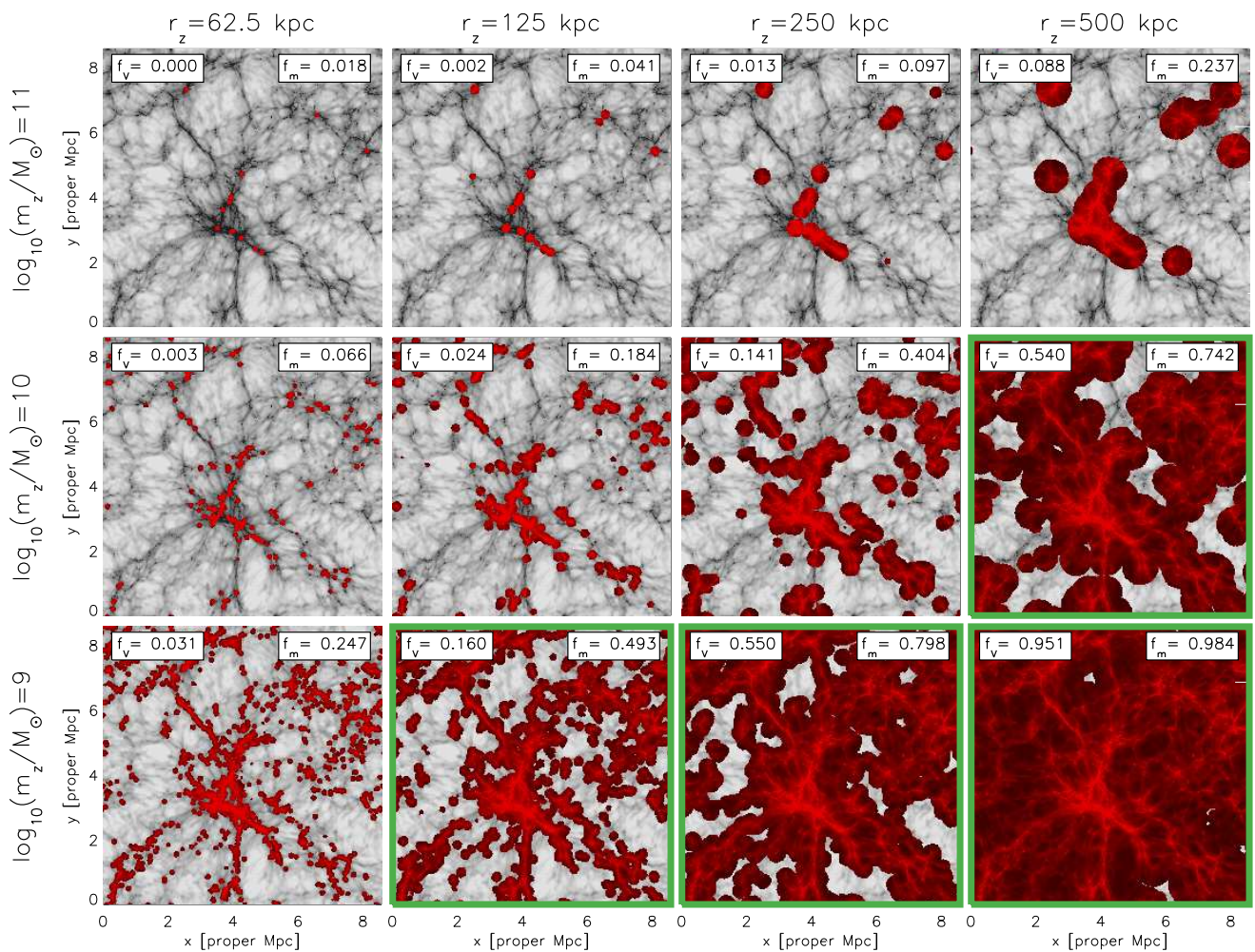}
\end{center}
\caption{Thin (1 comoving Mpc/$h$ thick) slices of the $z=3$ gas
  density field through the centre of the \emph{L025N512} simulation
  for models with different $\renrich$ (the proper radius out to which
  galaxies enrich the IGM) and $\menrich$ (the minimum halo mass that
  is responsible for enriching the IGM).  Each density field is shown
  in grey-scale in regions where, for a given combination of
  $\renrich$ and $\menrich$, the gas is unenriched, and in red where
  metals are present.  Each row of plots corresponds to a different
  value of $\menrich$ and each column shows a different value of
  $\renrich$. The numbers at the top of each panel show what fraction
  of the total simulation volume ($\fvol$) and gas mass ($\fmass$) are
  enriched with metals in each model. Panels outlined in green show
  models that satisfy the constraints provided by the observed
  $\tauciv-\tauhi$ relation. All models that reproduce the
  observations have volume filling factors $>10\%$.}
\label{fig:density}
\end{figure*}

\begin{figure*}
\begin{center}
\includegraphics[width=\textwidth,clip]{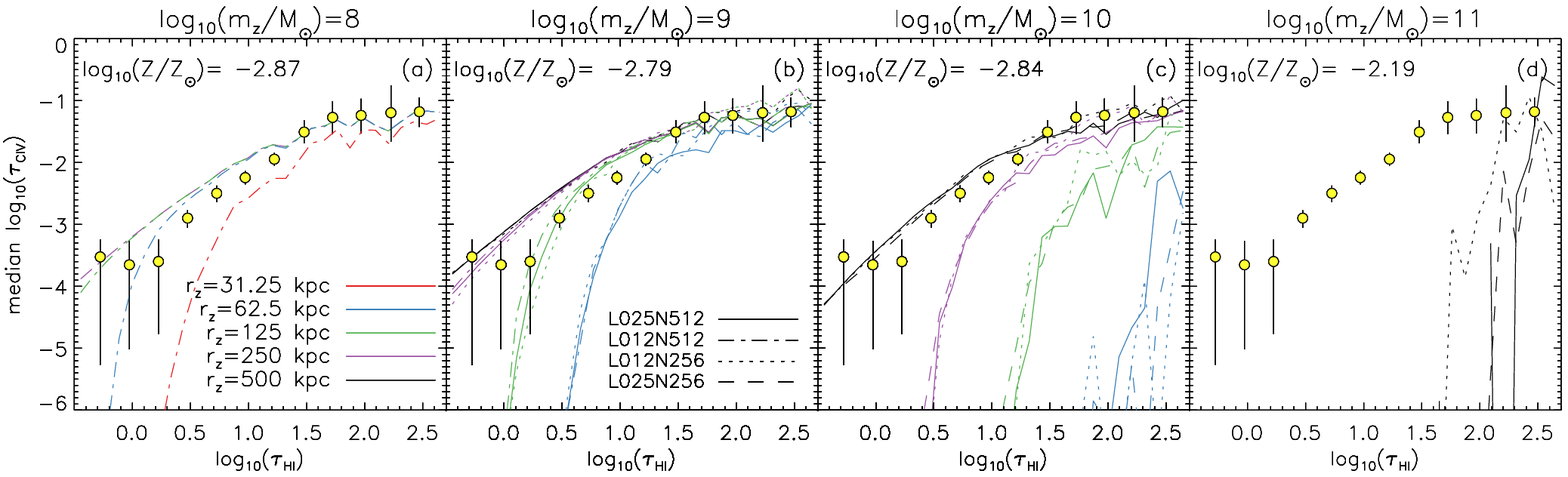}
\end{center}
\caption{The dependence of the relation between $\tauhi$ and $\tauciv$
  on the model parameters $\renrich$ (the physical radius out to which
  galaxies enrich the IGM) and $\menrich$ (the minimum halo mass that
  is capable of enriching the IGM).  Each panel shows a different
  choice for $\menrich$ and in each of the panels the yellow points
  with error bars show the observed relation of \citet{agui05}, with
  arrows representing upper limits. The curves show the simulation
  results, with the different colours representing the different
  choices for $\renrich$, and different line styles corresponding to
  different simulations.  The metallicity, $\zenrich$, required to
  match the normalisation of the observed data at the maximum value of
  $\tauhi$ (for the maximum value of $\renrich$) is shown at the top
  of each panel. The black curve in panel \emph{a} is not visible
  because for $\menrich=10^8\,\msun$ both $\renrich=250$\,kpc and
  $\renrich=500$\,kpc have $\fvol=1.00$ and hence predict identical
  results. All coloured curves in each panel are scaled to the same
  metallicity.  In order to match the observed median optical depth of
  $\civ$ in regions with a low $\hi$ optical depth, it is necessary
  that low-mass haloes ($\menrich\le10^{10.5}\,\msun$) enrich the IGM
  out to distances $>10^2\,$kpc.}

\label{fig:od}
\end{figure*}

Fig.~\ref{fig:density} shows a thin (1 comoving Mpc/$h$ thick) slice
of the density field at $z=3$ through the centre of the
\emph{L025N512} simulation.  Each panel shows a different combination
of $\renrich$ and $\menrich$. Gas with zero metallicity is shown in
grey-scale, while metal-enriched gas is coloured red.  In addition,
the values for $\fvol$ and $\fmass$ are indicated for each model.
Panels showing metal distributions that are consistent with the
observations (see below) are outlined in green.

The fractions of the mass and volume that are enriched do not track
each other in a simple way.  The ratio $\fmass/\fvol$ is always $\ge
1$ because metals are only placed around collapsed structures and thus
preferentially in overdense regions.  For $\menrich=10^{10}\,\msun$,
changing $\renrich$ from 31.25\,kpc to 500\,kpc changes the ratio
$\fmass/\fvol$ from 22.7 to 1.3, as a larger value of $\renrich$ at
fixed $\menrich$ allows metals to disperse further out of the
high-density peaks.  Similarly, changing $\menrich$ at fixed
$\renrich=250$~kpc, we find that the ratio $\fmass/\fvol$ rises from
2.8 for $\menrich=10^{9}\,\msun$ to 7.5 for $\menrich=10^{11}\,\msun$,
because more massive haloes are preferentially located in higher
density environments.  Given that we understand how changing the
pattern of metal enrichment alters the relative mass and volume
filling factors, we now ask how this impacts the $\tauciv-\tauhi$
relation.

The curves in Fig.~\ref{fig:od} show the relation between $\tauciv$
and $\tauhi$ in the synthetic absorption spectra.  Each panel
corresponds to a different $\menrich$. The solid lines in panels
\emph{b}, \emph{c} and \emph{d}
($\log_{10}(\menrich/\msun)={9,10,11}$, respectively) show the
predicted $\tauciv-\tauhi$ relation from the \emph{L025N512}
simulation with various imposed metallicity distributions. The
dot-dashed lines in panels \emph{a} and \emph{b}
($\log_{10}(\menrich/\msun)={8,9}$, respectively) show the predicted
$\tauciv-\tauhi$ relations for the small volume, high-resolution
simulation, \emph{L012N512}.  In every panel we compare our simulated
predictions to the observed optical depth pixel statistics of
\citet{scha03}, for the redshift range $2.479 \le z \le 4.033$, as
published in \citet{agui05}, shown as the yellow points with $1\sigma$
error bars.  The data come from six quasar spectra,
Q0420\textendash388, Q1425+604, Q2126\textendash158, Q1422+230,
Q0055\textendash269, and Q1055+461, that were taken with either the
Keck/HIRES or the VLT/UVES.  A full description of the sample is given
in \citet{scha03}.  In each panel $\zenrich$ was chosen such that the
$\renrich=500$~kpc curve exactly matches the observations at the
highest value of $\tauhi$ and the metallicity required for this
normalisation is given in each panel.

In the present work we are not aiming to reproduce the shape of the
$\tauciv-\tauhi$ relation in detail and, indeed, would not necessarily
expect our simple models to be capable of doing this.  Rather, we
require that the models predict a median $\tauciv-\tauhi$ relation
that is consistent with, or larger, than observed and that the
$\tauciv-\tauhi$ relation is not steeper than observed. We effectively
combine the two conditions by scaling $\zenrich$ in each run to match
the normalisation of the observed $\tauciv-\tauhi$ relation at
$\log_{10}\tauhi=2.5$, the largest optical depth probed by the
observations, and then requiring the model to predict median $\tauciv$
at low $\tauhi$ that are consistent with or greater than observed.

We stress that overprediction of $\tauciv$ is not a problem at small
$\tauhi$ because our simple models make the assumption that the
metallicity inside each enriched bubble is constant with radius.  The
overestimate at low $\tauciv$ could therefore be solved by imposing a
metallicity
that decreases with radius.  On the other hand, we will assume that
underprediction of $\tauciv$ at low $\tauhi$ signals the failure of the model because
a metallicity that increases with radius is likely unphysical. Such an unphysical metallicity gradient would also have been required if we had chosen to scale the metallicity to match the low $\tauhi$ points, because the unsuccessful, constant metallicity models predict much steeper $\tauciv-\tauhi$ relations than observed (see Fig.~\ref{fig:od}). In fact, such an approach would not be possible for most of the models that we rule out, because they typically predict the median $\tauciv$ to be zero at low $\tauhi$.

In principle, we could produce
significantly better fits to the $\tauciv-\tauhi$ relation by
considering more complex models (e.g.\ including metallicity gradients, anisotropic outflows or scatter in metallicities), but that is not the aim of the present study. Given that such models would still not be self-consistent if the enrichment is done in post-processing, it is not obvious that the use of more complex models would be justified.

Before proceeding, we evaluate the effect of our simulation's finite
box size and numerical resolution on our results.  Firstly, the dotted
curves in panels \emph{b}, \emph{c} and \emph{d} show the effect of
decreasing the box size by a factor of two in each dimension, while
keeping the numerical resolution fixed (\emph{L012N256}
vs. \emph{L025N512}).  It is clear that for
$\menrich\le10^{10}\,\msun$ our results are converged with respect to
box size, but that for $\menrich=10^{12}\,\msun$ the 12.5\,Mpc/$h$ and
25\,Mpc/$h$ volumes predict significantly different $\tauciv-\tauhi$
relations, indicating that for haloes of this mass the simulation
results from the 25\,Mpc/$h$ volume are not necessarily converged.
Note, however, that allowing only haloes with masses $\ge
10^{11}\,\msun$ to enrich the IGM yields $\tauciv-\tauhi$ relations
that are far steeper than those allowed by the observations. We thus
conclude that our simulation boxes are sufficiently large for our
purposes.  Secondly, the long-dashed curves in panels \emph{c} and
\emph{d} show the effect of degrading the simulation mass resolution
by a factor of eight while keeping the box size constant
(\emph{L025N256} vs. \emph{L025N512}). Decreasing the resolution does
not significantly change any of the conclusions derived from this
analysis.

Considering first the case where only haloes with mass
$>10^{11}\,\msun$ enrich the IGM, we find that even if such objects
were able to enrich out to a radius of $500$~kpc, they would fall far
short of being able to account for the observed $\tauciv$ at
$\tauhi<10$ because there are too few of them to enrich enough of the
low-density gas.  If we now consider a lower value of
$\menrich=10^{9}\,\msun$, the lower clustering strength and higher
number density of these less massive haloes allows them to pollute
lower density gas, even if they enrich out to much smaller distances.
All values of $\renrich\ge 125$~kpc provide a good match to the
observations (note that measurements of
$\tauciv$ are upper limits for $\log_{10}(\tauhi) < 0.3$).  The intermediate value of
$\menrich=10^{10}\,\msun$ provides, as expected, results that are
intermediate between the two cases presented above: the observations
are reproduced for $\renrich=500\,$kpc but not for $\renrich=250\,$kpc
and smaller.  Considering now the extremely high-resolution simulation
(\emph{L012N512}; panel \emph{a}), we find that if $\menrich$ is as
low as $10^8\,\msun$, then the observed $\tauciv-\tauhi$ relation can
be matched even if $\renrich$ is as small as 62.5\,kpc, but not for
$\renrich=$31.25\,kpc.

\begin{figure}
\begin{center}
\includegraphics[width=8.3cm,clip]{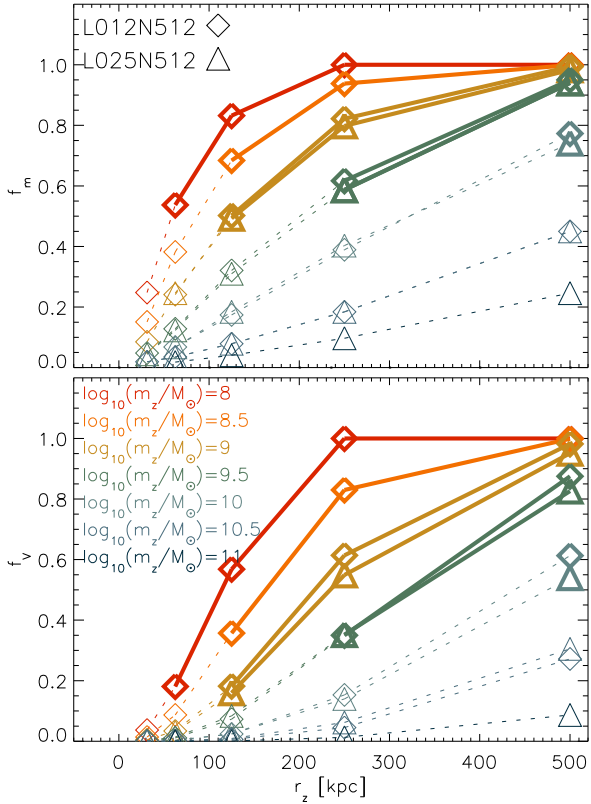}
\end{center}
\caption{Fraction of the volume ($\fvol$; top panel) and gas mass ($\fmass$; bottom panel) of
  the simulation that is enriched with metals as a function of
  $\renrich$ (the proper radius out to which galaxies enrich the
  IGM).  Each colour represents a different choice for $\menrich$ (the
  minimum halo mass that is capable of enriching the IGM), and the
  diamonds (triangles) show results for \emph{L012N512}
  (\emph{L025N512}).  Thick curves and symbols show models that
  reproduce the $\tauciv-\tauhi$ relation, thin, dotted lines show
  models that do not.  We require $\fvol>0.1$ and $\fmass>0.5$ in
  order to match the observations.}
\label{fig:rz-fv}
\end{figure}

Our results for the volume and mass filling factors of metals are
summarised in Fig.~\ref{fig:rz-fv}, where we show the mass (upper
panel) and volume (lower panel) filling fractions as a function of
$\renrich$ for different values of $\menrich$.  Models that predict
enough $\civ$ at low $\tauhi$ are indicated with bold symbols and are
connected with solid, thick lines.  It is clear that in order to
reproduce the observations, we require that the IGM is enriched
primarily by low-mass ($m_{\rm tot}<10^{10}\,\msun$) haloes that are
capable of driving gas out to distances $>10^2\,$kpc by $z=3$.
Allowing only haloes with masses $\ge10^{10.5}\,\msun$ to enrich the
IGM, results in metal distributions that fall far short of the
observations for all sensible values of $\renrich$. Models relying on
(the progenitors of) observed Lyman break galaxies \citep[$m_{\rm
    tot}\sim10^{12}\,\msun$;][]{adel05} to do the enrichment, predict
filling factors that are far too small to account for the observations
and have not even been plotted.  \emph{In our models, the volume filling factor of metals must be $>10\%$ in order to reproduce the observations}, indicating that the observed $\tauciv-\tauhi$
relation tells us robustly that a significant fraction of the volume
and mass of the IGM is enriched by metals. On the
other hand, a model volume filling factor $>10\%$ is not necessarily
sufficient.  Models that attain such large volume fractions by
enriching the gas around high-mass haloes to very large distances
often fail to reproduce the observations.  We note that the values of the volume and
mass filling fractions that we infer by comparing our models to the observations are approximate, even ignoring uncertainties associated with small-scale metal mixing. The filling fractions could change by factors of a few due to, for example, scatter in the metallicity around galaxies of a fixed mass and non-spherical bubbles.

\begin{figure}
\begin{center}
\includegraphics[width=8.3cm,clip]{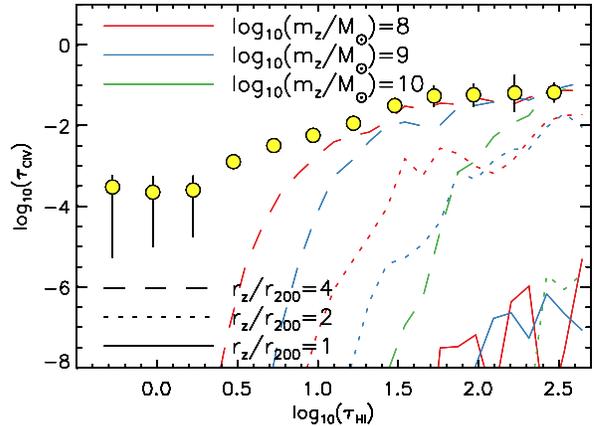}
\end{center}
\caption{The $z=3$ $\tauciv-\tauhi$ relation for models in which
  $\renrich$ (the physical radius out to which galaxies enrich the
  IGM) is set to a fixed multiple of the halo virial radius.  Yellow
  points with error bars represent the observations presented in
  Aguirre et al. (2005). Each curve represents a different imposed
  metal enrichment pattern, with different colours corresponding to
  different values of $\menrich$, and different line styles
  representing different choices for $\renrich$, as indicated in the
  legends. Even if all haloes can enrich the IGM out to four times
  their own virial radius, the observed $\tauciv-\tauhi$ relation is
  not reproduced. This implies that dynamical effects alone are
  insufficient to explain the observed distribution of $\civ$.}
\label{fig:rvir}
\end{figure}

Finally, in Fig.~\ref{fig:rvir} we show the effect of allowing haloes
to enrich out a fixed multiple of their virial radii, $r_{200}$.  The
yellow data points with error bars show again the observations of
\citet{scha03} as presented in \citet{agui05}, and the curves show the
results for the imposed metallicity distributions indicated in the
legend.  It is striking that even if all haloes that form stars are
able to pollute the IGM as far out as four times their own virial
radius, the low optical depth part of the $\tauciv-\tauhi$ relation is
not reproduced.  As dynamical processes (e.g.\ tidal or ram pressure stripping) are only expected to affect the metal distribution on scales $\la r_{200}$, this implies that some sort of ejective feedback is necessary in order to pollute the low-density IGM. The conclusion that dynamical processes such as stripping are not, on their own, sufficient is consistent with the models of \citet{agui01b} as well as with the hydrodynamical simulations of \citet{wier11}. 

\section{Discussion and conclusions}
\label{sec:conclusions}

By combining realistic cosmological density and temperature
distributions with toy models for the metal distribution, in which only
haloes more massive than a critical mass ($\menrich$) are able to
enrich a spherical region (of proper radius $\renrich$) to a metallicity of
$\zenrich$, we have investigated which metal distributions can
reproduce the observations of $\civ$ associated with weak $\hi$ absorption ($\tau_{\rm HI} \ll 10$) as measured in quasar
absorption spectra.

The results presented in Sec.~\ref{sec:results} imply that in order to
match the observed median optical depth of $\civ$ as a function of
$\tau_{\rm HI}$, we require that low-mass haloes, $m_{200}<
10^{10}\,\msun$, are able to drive metals into the IGM, enriching the
gas around them out to proper distances of $\ga 10^2$~kpc to
metallicities of $\sim10^{-3}\,\zsun$.

We now verify that this scenario is physically possible by comparing
the required metal mass to an estimate of the maximum allowed mass of
carbon in the Universe.  We can estimate the maximum allowed mass of
carbon in the IGM from the stellar mass density at $z=3$.  Under the
assumption that all of the stars formed in a single burst at $t=0$,
and assuming a Chabrier stellar initial mass function and using the
lifetimes, yields and supernova rates used in the simulation and
summarised in \citet{wier09}, the maximum allowed cosmic density of
carbon is $0.006\rho_\ast$, where $\rho_\ast$ is the cosmic density in
stars and has been measured to be $\rho_\ast=10^{7.55}\,\msun/{\rm
  Mpc}^3$ \citep{marc09} at $z=3$, giving us a maximum allowed total
density in carbon of $\rho_{{\rm C}}=2.2\times10^5\,\msun/{\rm
  Mpc}^3$. For the models that can successfully reproduce the observed
optical depth distributions, the cosmic density of carbon varies from
$8\times10^{3}\,\msun/{\rm Mpc}^3$ ($\menrich=10^{10}\,\msun$;
$\renrich=250\,{\rm kpc}$) to $2\times10^{4}\,\msun/{\rm Mpc}^3$
($\menrich=10^{9}\,\msun$; $\renrich=500\,{\rm kpc}$).  In each case
the total amount of carbon is an order of magnitude below the maximum
allowed amount\footnote{Including a lognormal scatter in the metal
  optical depth at the level measured by \citet{scha03} would raise the
  metallicity by a factor of a few, which is still comfortably within
  the allowed limits.}, so the models are physically reasonable on
these grounds.

We thus conclude that in order to recover the observed median $\civ$
optical depth in regions of low $\hi$ optical depth
($\tauhi\sim1-10$), we require that the galaxies in low-mass haloes
($\menrich<10^{10}\,\msun$) enrich the IGM out to distances $\ga
10^2\,$kpc.  Galaxies residing in much higher mass haloes
($\menrich>10^{10.5}\,\msun$) are too rare and too strongly clustered
to contribute significantly to the enrichment of the low-density
IGM. In every one of the models that is capable of reproducing the
observations, the metal volume filling factor is $>10\%$ and the gas
mass fraction enriched with metals is $>50\%$.
 
As discussed in detail in section~\ref{sec:metals}, our models are not
fully self-consistent in the sense that we have imposed metal
distribution is post-processing. Although we found that
post-processing hydro simulations without winds or with very strong
winds led to identical conclusions, we cannot rule out the possitibity
that future self-consistent models would yield different
results. Unfortunately, current simulations suffer from large
uncertainties due to their use of subgrid recipes for the generation
of winds and still lack the resolution to resolve outflows from
low-mass galaxies in a representative volume.

Two additional caveats must, however, be stressed. Firstly, the models
presented here implicitly assume that the metals are well mixed on
small scales. If, as suggested by observations \citep[e.g.][]{scha07},
the intergalactic metals are concentrated in metal-rich patches which
together account for large covering factors, then the required filling
factors could be smaller. In that case the filling factors derived
here apply to the metal distribution smoothed over the scales that are
somewhat smaller than the size of the bubbles, i.e.\ tens of kpc.

Secondly, there exists considerable uncertainty in the spectral shape
of the ionizing background, which leads to uncertainties the fraction
of carbon that exists as $\civ$. \citet{agui08} considered several
models for the UV background, including some extreme ones, and found
that only the fiducial \citet{haar01} model, which is the model used
here, resulted in reasonable values for the relative abundances
inferred from observations. \citet{scha03} found that assuming a much
harder (softer) spectrum would flatten (steepen) the
metallicity-density relation inferred from the observed
$\tauciv$-$\tauhi$ relation, which implies that our models would
predict steeper (flatter) $\tauciv$-$\tauhi$ relations than for our
standard UV background, which would increase (decrease) the inferred
filling factors.

Our finding that the observations imply that the IGM was enriched by
very low-mass galaxies is in agreement with a variety of theoretical
studies
\citep[e.g.][]{agui01a,mada01,scan02,thac02,samu08,oppe09,wier10}. For
example, by modelling the propagation of galactic winds in already
completed simulations, \citet{agui01a} found that in order to explain
the metallicities measured in the low column density part of the IGM,
galaxies with baryonic masses $>10^{8.5}\,\msun$ needed to launch
winds with velocities of at least $200-300$~km/s.  The results
presented here are also consistent with \citet{wier10}, who used fully
self-consistent hydrodynamical simulations (including the simulations
underlying our models) that massive haloes ($>10^{12}\,\msun$) are
unimportant for the enrichment of the diffuse IGM and that most of the
metals that reside in the IGM at $z=0$ were ejected by a population of
low-mass galaxies at high redshift.

In contrast, \citet{scan06} found that in order to match the strong
clustering of $\civ$ lines that they measured in their observations,
the absorption needed to be generated primarily by gas that is
strongly clustered around massive galaxies.  Their best fit model
required $\menrich=10^{12}\,\msun$ and $\renrich=500\,{\rm kpc}$.
Using these parameters in our model leads to such low volume
($\fvol=6\times10^{-3}$) and mass ($\fmass=0.03$) filling factors that
the corresponding curves would not appear on any of the plots
presented here.  We plan to explore the apparent tension between line
clustering and optical depth statistics in future work.

In summary, we have found that in order for a simulated cosmological
gas density field to reproduce the observed $\tauciv-\tauhi$ relation,
we require that both the fractions of the volume and mass of the IGM
that have been polluted with metals are substantial, with all our
successful parameter choices yielding model volume filling factors
greater than 10\% and model mass filling factors greater than 50\%.
The models favour metals being ejected from a population of low-mass
($m_{\rm tot}<10^{10}\,\msun$) haloes at high redshift.

\section*{Acknowledgments}
The authors would like to thank all the members of the OWLS team for
useful discussions; Anthony Aguirre, Ben Oppenheimer, and Olivera
Rakic for a careful reading of the manuscript; and Rob Wiersma for
both a reading of the manuscript and help with the calculation of
stellar yields. We are also grateful to the anonymous referee for a
constructive report. The simulations employed in this study were run
on Stella, the LOFAR BlueGene/L system in Groningen and on the
Cosmology Machine at the Institute for Computational Cosmology in
Durham as part of the Virgo Consortium research programme. This work
was sponsored by National Computing Facilities Foundation (NCF) for
the use of supercomputer facilities, with financial support from the
Netherlands Organization for Scientific Research (NWO), also through
an NWO Vidi grant.

\bibliographystyle{mn2e}

\begin{thebibliography}{67}
\expandafter\ifx\csname natexlab\endcsname\relax\def\natexlab#1{#1}\fi

\bibitem[{{Adelberger} {et~al.}(2005){Adelberger}, {Steidel}, {Pettini},
  {Shapley}, {Reddy}, \& {Erb}}]{adel05}
{Adelberger} K.~L., {Steidel} C.~C., {Pettini} M., {Shapley} A.~E., {Reddy}
  N.~A., {Erb} D.~K., 2005, ApJ, 619, 697

\bibitem[{{Aguirre} {et~al.}(2008){Aguirre}, {Dow-Hygelund}, {Schaye}, \&
  {Theuns}}]{agui08}
{Aguirre} A., {Dow-Hygelund} C., {Schaye} J., {Theuns} T., 2008, ApJ, 689, 851

\bibitem[{{Aguirre} {et~al.}(2001{\natexlab{a}}){Aguirre}, {Hernquist},
  {Schaye}, {Katz}, {Weinberg}, \& {Gardner}}]{agui01b}
{Aguirre} A., {Hernquist} L., {Schaye} J., {Katz} N., {Weinberg} D.~H.,
  {Gardner} J., 2001{\natexlab{a}}, ApJ, 561, 521

\bibitem[{{Aguirre} {et~al.}(2001{\natexlab{b}}){Aguirre}, {Hernquist},
  {Schaye}, {Weinberg}, {Katz}, \& {Gardner}}]{agui01a}
{Aguirre} A., {Hernquist} L., {Schaye} J., {Weinberg} D.~H., {Katz} N.,
  {Gardner} J., 2001{\natexlab{b}}, ApJ, 560, 599

\bibitem[{{Aguirre} {et~al.}(2005){Aguirre}, {Schaye}, {Hernquist}, {Kay},
  {Springel}, \& {Theuns}}]{agui05}
{Aguirre} A., {Schaye} J., {Hernquist} L., {Kay} S., {Springel} V., {Theuns}
  T., 2005, ApJL, 620, L13

\bibitem[{{Aracil} {et~al.}(2004){Aracil}, {Petitjean}, {Pichon}, \&
  {Bergeron}}]{arac04}
{Aracil} B., {Petitjean} P., {Pichon} C., {Bergeron} J., 2004, A\&A, 419, 811

\bibitem[{{Bertone} {et~al.}(2005){Bertone}, {Stoehr}, \& {White}}]{bert05}
{Bertone} S., {Stoehr} F., {White} S.~D.~M., 2005, MNRAS, 359, 1201

\bibitem[{{Carswell} {et~al.}(2002){Carswell}, {Schaye}, \& {Kim}}]{cars02}
{Carswell} B., {Schaye} J., {Kim} T., 2002, ApJ, 578, 43

\bibitem[{{Cen} \& {Chisari}(2011)}]{cen10}
{Cen} R., {Chisari} N.~E., 2011, ApJ, 731, 11

\bibitem[{{Cen} {et~al.}(2005){Cen}, {Nagamine}, \& {Ostriker}}]{cen05}
{Cen} R., {Nagamine} K., {Ostriker} J.~P., 2005, ApJ, 635, 86

\bibitem[{{Cen} \& {Ostriker}(2006)}]{cen06}
{Cen} R., {Ostriker} J.~P., 2006, ApJ, 650, 560

\bibitem[{{Cowie} {et~al.}(1995){Cowie}, {Hu}, \& {Songaila}}]{cowi95}
{Cowie} L.~L., {Hu} E.~M., {Songaila} A., 1995, AJ, 110, 1576

\bibitem[{{Cowie} \& {Songaila}(1998)}]{cowi98}
{Cowie} L.~L., {Songaila} A., 1998, Nature, 394, 44

\bibitem[{{Dalla Vecchia} \& {Schaye}(2008)}]{dall08}
{Dalla Vecchia} C., {Schaye} J., 2008, MNRAS, 387, 1431

\bibitem[{{Efstathiou}(1992)}]{efst92}
{Efstathiou} G., 1992, MNRAS, 256, 43P

\bibitem[{{Ellison} {et~al.}(2000){Ellison}, {Songaila}, {Schaye}, \&
  {Pettini}}]{elli00}
{Ellison} S.~L., {Songaila} A., {Schaye} J., {Pettini} M., 2000, AJ, 120, 1175

\bibitem[{{Faucher-Gigu{\`e}re} {et~al.}(2008){Faucher-Gigu{\`e}re}, {Lidz},
  {Hernquist}, \& {Zaldarriaga}}]{fauc08}
{Faucher-Gigu{\`e}re} C., {Lidz} A., {Hernquist} L., {Zaldarriaga} M., 2008,
  ApJ, 688, 85

\bibitem[{{Ferland} {et~al.}(1998){Ferland}, {Korista}, {Verner}, {Ferguson},
  {Kingdon}, \& {Verner}}]{ferl98}
{Ferland} G.~J., {Korista} K.~T., {Verner} D.~A., {Ferguson} J.~W., {Kingdon}
  J.~B., {Verner} E.~M., 1998, PASP, 110, 761

\bibitem[{{Germain} {et~al.}(2009){Germain}, {Barai}, \& {Martel}}]{germ09}
{Germain} J., {Barai} P., {Martel} H., 2009, ApJ, 704, 1002

\bibitem[{{Haardt} \& {Madau}(2001)}]{haar01}
{Haardt} F., {Madau} P., 2001, in Clusters of Galaxies and the High Redshift
  Universe Observed in X-rays, {D.~M.~Neumann \& J.~T.~V.~Tran}, ed.

\bibitem[{{Kobayashi} {et~al.}(2007){Kobayashi}, {Springel}, \&
  {White}}]{koba07}
{Kobayashi} C., {Springel} V., {White} S.~D.~M., 2007, MNRAS, 376, 1465

\bibitem[{{Madau} {et~al.}(2001){Madau}, {Ferrara}, \& {Rees}}]{mada01}
{Madau} P., {Ferrara} A., {Rees} M.~J., 2001, ApJ, 555, 92

\bibitem[{{Marchesini} {et~al.}(2009){Marchesini}, {van Dokkum}, {F{\"o}rster
  Schreiber}, {Franx}, {Labb{\'e}}, \& {Wuyts}}]{marc09}
{Marchesini} D., {van Dokkum} P.~G., {F{\"o}rster Schreiber} N.~M., {Franx} M.,
  {Labb{\'e}} I., {Wuyts} S., 2009, ApJ, 701, 1765

\bibitem[{{Martin}(2005)}]{mart05}
{Martin} C.~L., 2005, ApJ, 621, 227

\bibitem[{{Oppenheimer} \& {Dav{\'e}}(2006)}]{oppe06}
{Oppenheimer} B.~D., {Dav{\'e}} R., 2006, MNRAS, 373, 1265

\bibitem[{{Oppenheimer} \& {Dav{\'e}}(2008)}]{oppe08}
---, 2008, MNRAS, 387, 577

\bibitem[{{Oppenheimer} {et~al.}(2009){Oppenheimer}, {Dav{\'e}}, \&
  {Finlator}}]{oppe09}
{Oppenheimer} B.~D., {Dav{\'e}} R., {Finlator} K., 2009, MNRAS, 396, 729

\bibitem[{{Pettini} {et~al.}(2001){Pettini}, {Shapley}, {Steidel}, {Cuby},
  {Dickinson}, {Moorwood}, {Adelberger}, \& {Giavalisco}}]{pett01}
{Pettini} M., {Shapley} A.~E., {Steidel} C.~C., {Cuby} J., {Dickinson} M.,
  {Moorwood} A.~F.~M., {Adelberger} K.~L., {Giavalisco} M., 2001, ApJ, 554, 981

\bibitem[{{Pieri} {et~al.}(2010){Pieri}, {Frank}, {Weinberg}, {Mathur}, \&
  {York}}]{pier10}
{Pieri} M.~M., {Frank} S., {Weinberg} D.~H., {Mathur} S., {York} D.~G., 2010,
  ApJL, 724, L69

\bibitem[{{Pieri} \& {Haehnelt}(2004)}]{pier04}
{Pieri} M.~M., {Haehnelt} M.~G., 2004, MNRAS, 347, 985

\bibitem[{{Pieri} {et~al.}(2007){Pieri}, {Martel}, \& {Grenon}}]{pier07}
{Pieri} M.~M., {Martel} H., {Grenon} C., 2007, ApJ, 658, 36

\bibitem[{{Pinsonneault} {et~al.}(2010){Pinsonneault}, {Martel}, \&
  {Pieri}}]{pins10}
{Pinsonneault} S., {Martel} H., {Pieri} M.~M., 2010, ApJ, 725, 2087

\bibitem[{{Quinn} {et~al.}(1996){Quinn}, {Katz}, \& {Efstathiou}}]{quin96}
{Quinn} T., {Katz} N., {Efstathiou} G., 1996, MNRAS, 278, L49

\bibitem[{{Rakic} {et~al.}(2011){Rakic}, {Schaye}, {Steidel}, \&
  {Rudie}}]{raki10}
{Rakic} O., {Schaye} J., {Steidel} C.~C., {Rudie} G.~C., 2011, MNRAS, 414, 3265

\bibitem[{{Rauch} {et~al.}(1997){Rauch}, {Miralda-Escude}, {Sargent}, {Barlow},
  {Weinberg}, {Hernquist}, {Katz}, {Cen}, \& {Ostriker}}]{rauc97}
{Rauch} M., {Miralda-Escude} J., {Sargent} W.~L.~W., {Barlow} T.~A., {Weinberg}
  D.~H., {Hernquist} L., {Katz} N., {Cen} R., {Ostriker} J.~P., 1997, ApJ, 489,
  7

\bibitem[{{Samui} {et~al.}(2008){Samui}, {Subramanian}, \& {Srianand}}]{samu08}
{Samui} S., {Subramanian} K., {Srianand} R., 2008, MNRAS, 385, 783

\bibitem[{{Scannapieco} {et~al.}(2002){Scannapieco}, {Ferrara}, \&
  {Madau}}]{scan02}
{Scannapieco} E., {Ferrara} A., {Madau} P., 2002, ApJ, 574, 590

\bibitem[{{Scannapieco} {et~al.}(2006){Scannapieco}, {Pichon}, {Aracil},
  {Petitjean}, {Thacker}, {Pogosyan}, {Bergeron}, \& {Couchman}}]{scan06}
{Scannapieco} E., {Pichon} C., {Aracil} B., {Petitjean} P., {Thacker} R.~J.,
  {Pogosyan} D., {Bergeron} J., {Couchman} H.~M.~P., 2006, MNRAS, 365, 615

\bibitem[{{Schaye}(2001)}]{scha01}
{Schaye} J., 2001, ApJ, 559, 507

\bibitem[{{Schaye} \& {Aguirre}(2005)}]{scha05}
{Schaye} J., {Aguirre} A., 2005, in IAU Symposium, Vol. 228, From Lithium to
  Uranium: Elemental Tracers of Early Cosmic Evolution, {V.~Hill, P.~Fran{\c
  c}ois, \& F.~Primas}, ed., pp. 557--568

\bibitem[{{Schaye} {et~al.}(2003){Schaye}, {Aguirre}, {Kim}, {Theuns}, {Rauch},
  \& {Sargent}}]{scha03}
{Schaye} J., {Aguirre} A., {Kim} T., {Theuns} T., {Rauch} M., {Sargent}
  W.~L.~W., 2003, ApJ, 596, 768

\bibitem[{{Schaye} {et~al.}(2007){Schaye}, {Carswell}, \& {Kim}}]{scha07}
{Schaye} J., {Carswell} R.~F., {Kim} T., 2007, MNRAS, 379, 1169

\bibitem[{{Schaye} \& {Dalla Vecchia}(2008)}]{scha08}
{Schaye} J., {Dalla Vecchia} C., 2008, MNRAS, 383, 1210

\bibitem[{{Schaye} {et~al.}(2010){Schaye}, {Dalla Vecchia}, {Booth}, {Wiersma},
  {Theuns}, {Haas}, {Bertone}, {Duffy}, {McCarthy}, \& {van de Voort}}]{scha10}
{Schaye} J., {Dalla Vecchia} C., {Booth} C.~M., {Wiersma} R.~P.~C., {Theuns}
  T., {Haas} M.~R., {Bertone} S., {Duffy} A.~R., {McCarthy} I.~G., {van de
  Voort} F., 2010, MNRAS, 402, 1536

\bibitem[{{Schaye} {et~al.}(2000{\natexlab{a}}){Schaye}, {Rauch}, {Sargent}, \&
  {Kim}}]{scha00}
{Schaye} J., {Rauch} M., {Sargent} W.~L.~W., {Kim} T., 2000{\natexlab{a}},
  ApJL, 541, L1

\bibitem[{{Schaye} {et~al.}(2000{\natexlab{b}}){Schaye}, {Theuns}, {Rauch},
  {Efstathiou}, \& {Sargent}}]{scha00temp}
{Schaye} J., {Theuns} T., {Rauch} M., {Efstathiou} G., {Sargent} W.~L.~W.,
  2000{\natexlab{b}}, MNRAS, 318, 817

\bibitem[{{Schwartz} \& {Martin}(2004)}]{schw04}
{Schwartz} C.~M., {Martin} C.~L., 2004, ApJ, 610, 201

\bibitem[{{Shapley} {et~al.}(2003){Shapley}, {Steidel}, {Pettini}, \&
  {Adelberger}}]{shap03}
{Shapley} A.~E., {Steidel} C.~C., {Pettini} M., {Adelberger} K.~L., 2003, ApJ,
  588, 65

\bibitem[{{Shen} {et~al.}(2010){Shen}, {Wadsley}, \& {Stinson}}]{shen10}
{Shen} S., {Wadsley} J., {Stinson} G., 2010, MNRAS, 407, 1581

\bibitem[{{Simcoe} {et~al.}(2004){Simcoe}, {Sargent}, \& {Rauch}}]{simc04}
{Simcoe} R.~A., {Sargent} W.~L.~W., {Rauch} M., 2004, ApJ, 606, 92

\bibitem[{{Songaila}(2005)}]{song05}
{Songaila} A., 2005, AJ, 130, 1996

\bibitem[{{Spergel} {et~al.}(2007){Spergel}, {Bean}, {Dor{\'e}}, \& {et
  al.}}]{sper07}
{Spergel} D.~N., {Bean} R., {Dor{\'e}} O., {et al.}, 2007, ApJS, 170, 377

\bibitem[{{Springel}(2005)}]{spri05b}
{Springel} V., 2005, MNRAS, 364, 1105

\bibitem[{{Springel} {et~al.}(2001){Springel}, {White}, {Tormen}, \&
  {Kauffmann}}]{spri01}
{Springel} V., {White} S.~D.~M., {Tormen} G., {Kauffmann} G., 2001, MNRAS, 328,
  726

\bibitem[{{Steidel} {et~al.}(2010){Steidel}, {Erb}, {Shapley}, {Pettini},
  {Reddy}, {Bogosavljevi{\'c}}, {Rudie}, \& {Rakic}}]{stei10}
{Steidel} C.~C., {Erb} D.~K., {Shapley} A.~E., {Pettini} M., {Reddy} N.,
  {Bogosavljevi{\'c}} M., {Rudie} G.~C., {Rakic} O., 2010, ApJ, 717, 289

\bibitem[{{Tescari} {et~al.}(2011){Tescari}, {Viel}, {D'Odorico}, {Cristiani},
  {Calura}, {Borgani}, \& {Tornatore}}]{tesc10}
{Tescari} E., {Viel} M., {D'Odorico} V., {Cristiani} S., {Calura} F., {Borgani}
  S., {Tornatore} L., 2011, MNRAS, 411, 826

\bibitem[{{Thacker} {et~al.}(2002){Thacker}, {Scannapieco}, \&
  {Davis}}]{thac02}
{Thacker} R.~J., {Scannapieco} E., {Davis} M., 2002, ApJ, 581, 836

\bibitem[{{Theuns} {et~al.}(1998){Theuns}, {Leonard}, {Efstathiou}, {Pearce},
  \& {Thomas}}]{theu98}
{Theuns} T., {Leonard} A., {Efstathiou} G., {Pearce} F.~R., {Thomas} P.~A.,
  1998, MNRAS, 301, 478

\bibitem[{{Theuns} {et~al.}(2002){Theuns}, {Viel}, {Kay}, {Schaye}, {Carswell},
  \& {Tzanavaris}}]{theu02}
{Theuns} T., {Viel} M., {Kay} S., {Schaye} J., {Carswell} R.~F., {Tzanavaris}
  P., 2002, ApJL, 578, L5

\bibitem[{{Thoul} \& {Weinberg}(1996)}]{thou96}
{Thoul} A.~A., {Weinberg} D.~H., 1996, ApJ, 465, 608

\bibitem[{{Tornatore} {et~al.}(2010){Tornatore}, {Borgani}, {Viel}, \&
  {Springel}}]{torn10}
{Tornatore} L., {Borgani} S., {Viel} M., {Springel} V., 2010, MNRAS, 402, 1911

\bibitem[{{Tytler} {et~al.}(1995){Tytler}, {Fan}, {Burles}, {Cottrell},
  {Davis}, {Kirkman}, \& {Zuo}}]{tytl95}
{Tytler} D., {Fan} X., {Burles} S., {Cottrell} L., {Davis} C., {Kirkman} D.,
  {Zuo} L., 1995, in QSO Absorption Lines, {G.~Meylan}, ed., pp. 289--+

\bibitem[{{Weinberg} {et~al.}(1997){Weinberg}, {Miralda-Escude}, {Hernquist},
  \& {Katz}}]{wein97}
{Weinberg} D.~H., {Miralda-Escude} J., {Hernquist} L., {Katz} N., 1997, ApJ,
  490, 564

\bibitem[{{Wiersma} {et~al.}(2010){Wiersma}, {Schaye}, {Dalla Vecchia},
  {Booth}, {Theuns}, \& {Aguirre}}]{wier10}
{Wiersma} R.~P.~C., {Schaye} J., {Dalla Vecchia} C., {Booth} C.~M., {Theuns}
  T., {Aguirre} A., 2010, MNRAS, 409, 132

\bibitem[{{Wiersma} {et~al.}(2009{\natexlab{a}}){Wiersma}, {Schaye}, \&
  {Smith}}]{wier08}
{Wiersma} R.~P.~C., {Schaye} J., {Smith} B.~D., 2009{\natexlab{a}}, MNRAS, 393,
  99

\bibitem[{{Wiersma} {et~al.}(2011){Wiersma}, {Schaye}, \& {Theuns}}]{wier11}
{Wiersma} R.~P.~C., {Schaye} J., {Theuns} T., 2011, MNRAS, 415, 353

\bibitem[{{Wiersma} {et~al.}(2009{\natexlab{b}}){Wiersma}, {Schaye}, {Theuns},
  {Dalla Vecchia}, \& {Tornatore}}]{wier09}
{Wiersma} R.~P.~C., {Schaye} J., {Theuns} T., {Dalla Vecchia} C., {Tornatore}
  L., 2009{\natexlab{b}}, MNRAS, 399, 574

\end{thebibliography}

\label{lastpage}

\end{document}